\documentstyle[aps,preprint,epsfig]{revtex}

\begin{document}
\draft
\title{Pathologies in the sticky limit of\\
hard-sphere-Yukawa models for colloidal fluids.\\
A possible correction.}
\author{Domenico Gazzillo \thanks{%
Author for correspondence. {\it E-mail address:} gazzillo@unive.it} and
Achille Giacometti}
\address{Istituto Nazionale per la Fisica della Materia and \\
Dipartimento di Chimica Fisica, Universit\`{a} di Venezia, \\
S. Marta DD 2137, I-30123 Venezia, Italy}
\date{\today }
\maketitle

\begin{abstract}
A known `sticky-hard-sphere' model, defined starting from a
hard-sphere-Yukawa potential and taking the limit of infinite amplitude and
vanishing range with their product remaining constant, is shown to be
ill-defined. This is because its Hamiltonian (which we call SHS2) leads to
an {\it exact }second virial coefficient which {\it diverges}, unlike that
of Baxter's original model (SHS1).

This deficiency has never been observed so far, since the
linearization implicit in the `mean spherical approximation' (MSA), within
which the model is analytically solvable, partly {\it masks} such a
pathology. To overcome this drawback and retain some useful features of
SHS2, we propose both a new model (SHS3) and a new closure (`modified MSA'),
whose combination yields an analytic solution formally identical with the
SHS2-MSA one. This mapping allows to recover many results derived from SHS2,
after a re-interpretation within a correct framework. Possible developments
are finally indicated.
\end{abstract}

\newpage

\section{INTRODUCTION}

In a seminal series of papers \cite{Baxter68,Baxter71a,Baxter71b} Baxter
first introduced the concept of the so-called `sticky hard sphere' (SHS)
models, as the simplest - albeit crude - modelization for real fluids of
spherical particles with a strong surface adhesion. In Baxter's original
formulation \cite{Baxter68,Baxter71a,Baxter71b,Barboy74} and its extension
to the multi-component case \cite{Perram75,Barboy79} (both hereafter
referred to as SHS1 model) the pair potential contains - in addition to a
hard sphere (HS) repulsion - a infinitely deep and narrow attractive
square-well, obtained according to a particular limiting procedure (Baxter's
`sticky limit') that keeps the second virial coefficient finite \cite%
{Baxter68}.

Although this model appears rather pathological at first sight, it includes
a number of interesting features which justify its wide popularity. First,
the Ornstein-Zernike (OZ) integral equation of the statistical-mechanical
theory of fluids can be analytically solved for it within the {\it %
Percus-Yevick} (PY) approximation and the solution exhibits a gas-liquid
transition \cite{Baxter68,Baxter71a,Baxter71b,Barboy74}. Second, Baxter's
model has already proven to be appropriate for describing some properties of
colloidal suspensions, micelles, microemulsions and protein solutions with
short-range interactions as well as some aspects of adsorption, flocculation
and percolation phenomena, solvent-mediated forces, ionic mixtures,
solutions with a small degree of size polydispersity and fluids of
chain-like molecules (for an illustrative, albeit not exhaustive, list of
references, see \cite{Stell91,Jamnik96,Santos98,Jamnik01}). All this means
that, in spite of its highly idealized character and known shortcomings \cite%
{Stell91}, the SHS1 model is able to capture some important physical
features of structure, thermodynamics, and phase behaviour of many real
systems.

However, the SHS1 model has its main drawback in its problematic application
to mixtures with a large number $p$ of components - as occurs for colloidal
suspensions with large polydispersity - since this case requires the
solution for a set of $p(p+1)/2$ coupled quadratic equations \cite{Perram75}%
, a task which cannot be accomplished analytically. This important fact has
originated a more recent attempt of finding an alternative SHS model, which
could be analytically tractable even in the general multi-component case %
\cite%
{Brey87,Mier89,Ginoza96,Tutschka98,Tutschka00,Tutschka01,Tutschka02,Gazzillo00,Gazzillo02a,Gazzillo02b,Gazzillo03,Timoneda89,Ginoza01}%
. For pure fluids, Brey {\it et al.} \cite{Brey87} proposed to start from a
HS-Yukawa (HSY) potential

\begin{equation}
\beta \phi _{{\rm HSY}}(r)=\left\{ 
\begin{array}{ccc}
+\infty , &  & 0<r<\sigma , \\ 
-K\ e^{-z(r-\sigma )}/r, &  & r\geq \sigma ,%
\end{array}%
\right.  \label{i1}
\end{equation}%
with 
\begin{equation}
K=zK_{0},\qquad K_{0}=\varepsilon _{0}^{\ast }\ \sigma ^{2},\text{\ \ \ \ \
\ \ }\varepsilon _{0}^{\ast }=\beta \varepsilon _{0}\equiv \frac{1}{%
12T^{\ast }},  \label{i2}
\end{equation}%
where $r$ is the distance between particles, $\beta =(k_{B}T)^{-1}$ ($k_{B}$
being Boltzmann's constant, and $T$ the temperature), $\sigma $ denotes the
HS diameter, $z$ the Yukawa inverse range, $\varepsilon _{0}$ an energy, and 
$T^{\ast }$ a reduced temperature (as in SHS1, the factor $12$ simplifies
subsequent analysis). The definition of this second Hamiltonian (SHS2 model
in the following) is completed by the definition of a `sticky limit', which
in this case amounts to taking $z\rightarrow +\infty $ \cite{Brey87}. It is
worth remarking that, unlike its counterpart in the SHS1 model, the starting
potential $\phi _{{\rm HSY}}(r)$ is{\it \ itself} independent of temperature.

For the SHS2 model (in both cases of pure fluids and mixtures), the OZ
equation can be solved analytically within the {\it mean spherical
approximation} (MSA) \cite{Mier89,Ginoza96}. It turns out that the SHS2-MSA
solution $q(r)$ (or $q_{ij}(r)$ ) for the Baxter form of the OZ equation has
exactly the same $r$-dependence as the SHS1-PY solution, i.e. both are
expressed in terms of a second-degree polynomial in $r$. However, the
difference between the two solutions lies in the density and temperature
dependence of their polynomial coefficients (see below), which makes the
SHS2-MSA solution readily usable even in the multi-component case \cite%
{Tutschka98,Tutschka00,Tutschka02,Gazzillo00,Gazzillo02a}, unlike the
SHS1-PY one.

Unfortunately, the common belief that the SHS1 and SHS2 Hamiltonians are
different but equivalent representations of a {\it unique} SHS potential %
\cite%
{Jamnik96,Jamnik01,Tutschka98,Tutschka00,Tutschka01,Tutschka02,Jamnik91,Hoye93}
as well as a failure in appreciating the subtle distinction between {\it %
model }and {\it solution} \cite{note0} have often generated a number of
misunderstandings and erroneous beliefs in the literature on SHS fluids. In
particular, both aforesaid PY and MSA solutions have sometimes been regarded
as corresponding to the same SHS model (Baxter's one) \cite%
{Jamnik96,Tutschka98,Tutschka00,Tutschka01,Tutschka02,Jamnik91}. On the
contrary, in the present paper we stress that the SHS1-PY and SHS2-MSA {\it %
solutions} are different not only as stemming from {\it different closures},
but, more importantly, since they refer to {\it different Hamiltonians}. In
fact, unlike SHS1, the SHS2 model {\it itself} is ill-defined from a
thermodynamical point of view, and thus SHS1 and SHS2 cannot be equivalent.
We will prove this point by considering the {\it exact} second virial
coefficient of the HSY potential (\ref{i1}) and showing that it {\it diverges%
} in the sticky limit (a very short, preliminary, account of these results
has been given in Ref. \cite{Gazzillo03}). As we further elaborate below,
this pathology is hidden in the SHS2-MSA compressibility equation of state
(EOS) but is mirrored by a similar singular behaviour of the corresponding
MSA virial and energy EOS's, never investigated in previous studies.

The second and main goal of the present paper is to propose a new model
(SHS3) which combines the advantages of both SHS1 and SHS2. As SHS1, it has
a finite second virial coefficient, and thus it is a well-defined model. As
SHS2 however, it admits a simple analytical solution within a new closure,
referred to as {\it modified mean spherical approximation }(mMSA) in the
following. The remarkable property of the SHS3-mMSA solution for $q(r)$ is that
it turns out to be formally identical with the SHS2-MSA one, and so are all
quantities which are immediately derivable from it (such as structural
properties and compressibility EOS). As a consequence of this mapping, all
SHS2-MSA results obtained in the past for these quantities can be recovered
after an appropriate re-interpretation. In addition, we will provide new
results for other quantities, notably the virial and energy EOS, for which
the SHS2-MSA solution badly fails. This is {\it not} the case in the
SHS3-mMSA solution, where the energy EOS turns out to be finite, albeit
mean-field-like. Finally, although the virial EOS displays a singular
behaviour within the SHS3-mMSA as well, we will argue that, unlike the SHS2
case, this divergence is a consequence of the deficiency of the mMSA
closure, and {\it not of the SHS3 model itself}.

Our findings thus provide a sound theoretical basis for a critical analysis
of the existing literature on SHS models, and allow to clarify
misunderstandings and discard incorrect results, previously reported in the
literature.

The plan of the remaining of the paper is the following. In Section II we
will briefly recall the main features of the SHS2-MSA solution and describe
the drawbacks of SHS2 in detail. The new model along with the new closure,
which constitute the central part of this work, are both discussed in
Section III. Final remarks and future perspectives close the paper (Section
IV).

\section{SHS2 MODEL\ AND\ ITS\ PATHOLOGIES}

For a one-component fluid of $N$ molecules in a volume $V,$ with spherically
symmetric interactions, the Baxter form of the OZ integral equation \cite%
{Baxter71a} is given by

\begin{equation}
\left\{ 
\begin{array}{l}
rc\left( r\right) =-q^{\prime }(r)+2\pi \rho \int_{r}^{\infty }dt\ q\left(
t-r\right) q^{\prime }\left( t\right) , \\ 
rh\left( r\right) =-q^{\prime }(r)+2\pi \rho \int_{0}^{\infty }dt\ q\left(
t\right) \left( r-t\right) h\left( |r-t|\right) ,%
\end{array}%
\right.  \label{ie2}
\end{equation}%
where $\rho =N/V$ denotes the number density, $c(r)$ the direct correlation
function (DCF), $h(r)=g(r)-1$, with $g(r)$ being the radial distribution
function (RDF). Moreover, the prime denotes differentiation with respect to $%
r$. Solving the Baxter equations is tantamount to determining the factor
correlation function $q(r)$, an auxiliary quantity from which $c\left(
r\right) $ and $h\left( r\right) $ can be easily derived.

An approximate integral equation can be obtained by adding to the OZ
equation some approximate `closure' relating $c\left( r\right) $, $h\left(
r\right) $ and the potential $\phi \left( r\right) $.

\subsection{MSA solution}

For potentials with a hard-core part, the {\it mean spherical approximation}
(MSA) reads

\begin{equation}
c_{{\rm MSA}}\left( r\right) =\left\{ 
\begin{array}{ccc}
-\left[ 1+\gamma \left( r\right) \right] , &  & 0<r<\sigma , \\ 
-\beta \phi _{{\rm tail}}\left( r\right) , &  & r\geq \sigma ,%
\end{array}%
\right.  \label{clos2}
\end{equation}%
where $\gamma \left( r\right) \equiv \rho \int d{\bf r}^{\prime }\ c\left(
r^{\prime }\right) h\left( |{\bf r-r}^{\prime }|\right) $, $\phi _{{\rm tail}%
}\left( r\right) $ is the potential outside the core. Since $g\left(
r\right) =1+\gamma \left( r\right) +c\left( r\right) ,$ the MSA may also be
written as%
\begin{equation}
g_{{\rm MSA}}\left( r\right) =e_{{\rm HS}}\left( r\right) \left[ 1+\gamma
\left( r\right) -\beta \phi _{{\rm tail}}\left( r\right) \right] ,
\label{clos5}
\end{equation}%
where $e_{{\rm HS}}\left( r\right) =\exp \left[ -\beta \phi _{{\rm HS}}(r)%
\right] =\theta \left( r-\sigma \right) ,$ with $\phi _{{\rm HS}}(r)$ being
the HS potential, and $\theta (x)$ the Heaviside step function ($\theta
(x)=0 $ for $x<0,$ and $\theta (x)=1$ for $x\geq 0$).

For the HSY potential the MSA closure becomes

\begin{equation}
c_{{\rm HSY-MSA}}(r)=K\ e^{-z(r-\sigma )}\ /r,\qquad r\ \geq \sigma ,
\label{d1}
\end{equation}%
where the parameters are the same as in Eq. (\ref{i1}). The analytic MSA
solution $q_{{\rm HSY-MSA}}(r),$ for the HSY fluid \cite{Hoye77} yields in
the sticky limit ($z\rightarrow +\infty ,$ ) the SHS2-MSA solution

\begin{equation}
q_{{\rm SHS2-MSA}}(r)=\left[ \frac{1}{2}a(r^{2}-\sigma ^{2})+b\sigma
(r-\sigma )+q_{\sigma }\sigma ^{2}\right] \theta \left( \sigma -r\right)
,\qquad \text{ for \ }r\geq 0,  \label{so1}
\end{equation}%
\begin{equation}
a=\frac{1+2\eta }{\left( 1-\eta \right) ^{2}}-\frac{12q_{\sigma }\eta }{%
1-\eta },\qquad b=-\frac{3\eta }{2\left( 1-\eta \right) ^{2}}+\frac{%
6q_{\sigma }\eta }{1-\eta },  \label{so2}
\end{equation}

\begin{equation}
q_{\sigma }=K_{0}/\sigma ^{2}=\text{\ }\varepsilon _{0}^{\ast }.  \label{so3}
\end{equation}%
where $\eta =(\pi /6)\rho \sigma ^{3}$ is the packing fraction. \noindent As
remarked, the expression ($\ref{so1}$)\ is formally identical with the
SHS1-PY solution with the crucial difference that $q_{\sigma }^{{\rm SHS2-MSA%
}}$ depends only on temperature (being proportional to $T^{\ast -1})$,
whereas $q_{\sigma }^{{\rm SHS1-PY}}$ depends on {\it both} temperature {\it %
and} density (in a more complex way) \cite{Baxter71a}.

Once $q(r)$ is known, all structural and thermodynamic properties can, in
principle, be calculated. It is worth mentioning that, with a few
exceptions\ \cite{Mier89,Tutschka01,Gazzillo02b,Timoneda89,Ginoza01,Hoye93},
the thermodynamic properties of SHS2 are still mostly unexplored. However,
we stress once again that the main aim of our paper is not to present new
results on the SHS2-MSA thermodynamics (although this task will be
accomplished too), but to reveal a dramatic fault of the SHS2 potential {\it %
itself} (irrespectively of any approximate closure), which, surprisingly
enough, has never been observed in the previous literature, but will emerge
from the following simple analysis.

\subsection{Sticky limit of the {\it exact} second virial coefficient of the
HSY fluid}

The Hamiltonian of the SHS1 model was introduced by Baxter as a limiting
case, through a clever definition of {\it both} a starting potential and a
limit procedure in such a way that the contribution of the vanishing
square-well tail to the second virial coeffient, $B_{2}$, remains finite and
nonzero. It is instructive to consider the behaviour of $B_{2}$ for the SHS2
model, by considering the result of the $z\rightarrow +\infty $ limit for
the {\it exact} second virial coefficient $B_{2\text{ }}^{{\rm HSY-exact}}$
of the HSY fluid. This can be calculated from the general definition, 
\begin{equation}
B_{2}=-2\pi \int_{0}^{\infty }dr\ r^{2}f(r),  \label{b2exact}
\end{equation}%
with $f(r)=e(r)-1,$ and $e(r)=\exp \left[ -\beta \phi \left( r\right) \right]
$ being the Boltzmann factor. One finds

\[
B_{2\text{ }}^{{\rm HSY-exact}}=\frac{2\pi }{3}\sigma ^{3}+\Delta B_{2\text{ 
}}^{{\rm HSY-exact}} 
\]%
where 
\[
\Delta B_{2\text{ }}^{{\rm HSY-exact}}=-2\pi \int_{\sigma }^{\infty }dr\
r^{2}\left\{ \exp \left[ zK_{0}\ e^{-z(r-\sigma )}/r\right] -1\right\} , 
\]%
It is now easy to show that $\Delta B_{2\text{ }}^{{\rm HSY-exact}}$
diverges in the $z\rightarrow +\infty $ limit. To this aim we note that, for 
$x=zK_{0}\ e^{-z(r-\sigma )}/r\geq 0$, one has $e^{x}-1\geq x+x^{2}/2$ and
hence we can use the bound

\begin{equation}
\int_{\sigma }^{\infty }dr\ r^{2}\left( e^{x}-1\right) \geq \int_{\sigma
}^{\infty }dr\ r^{2}\left( x+\frac{x^{2}}{2}\right) =K_{0}\left( \sigma +%
\frac{1}{z}\right) +\frac{K_{0}^{2}}{4}\ z.  \label{bound}
\end{equation}%
As the right hand side of Eq.(\ref{bound}) diverges as $z\rightarrow +\infty
,$ we have thus shown that 
\begin{equation}
B_{2\text{ }}^{{\rm SHS2-exact}}\equiv \lim_{z\rightarrow +\infty }B_{2\text{
}}^{{\rm HSY-exact}}=-\infty .  \label{d15}
\end{equation}%
This result is exact and {\it independent} of any closure, and reflects an
inconsistency of the HSY potential with the definition of sticky limit
employed when setting up the SHS2 Hamiltonian \cite{Brey87,Mier89}. As a
consequence, the SHS2 Hamiltonian is ill-defined from the outset, and the
corresponding model (which cannot be a different representation of Baxter's
one) {\it must be discarded}.

This is, however, a surprising result in some respect. One may rightfully
wonder why no trace of the pathological nature of the SHS2 model has ever
been revealed by a number of structural studies \cite%
{Tutschka98,Tutschka00,Tutschka02,Gazzillo00,Gazzillo02a} carried out on its
multi-component version. As discussed in Section IV, all these structural
results based upon the one and multi-component SHS2-MSA solutions are, in
fact, fully correct after their re-interpretation in terms of SHS3.
Before doing this, however, it is instructive to consider those SHS2-MSA
thermodynamic properties of the one-component fluid which have {\it not}
been investigated so far. This {\it will} display the pathological nature of
the model as shown next.

\subsection{Sticky limit of the MSA equations of state for the HSY fluid}

Mier-y-Teran {\it et al.} \cite{Mier89} obtained, for $Z=\beta P/\rho $, the
MSA{\it \ }compressibility{\it \ }$(C)${\it \ }EOS 
\begin{eqnarray}
Z_{C}^{{\rm SHS2-MSA}} &=&\frac{1+\eta +\eta ^{2}}{(1-\eta )^{3}}+\left[ 
\frac{4-7\eta }{(1-\eta )^{2}}+\frac{4}{\eta }\ln (1-\eta )\right] \frac{1}{%
T^{\ast }}  \nonumber \\
&&\ +\left[ \frac{2-\eta }{1-\eta }+\frac{2}{\eta }\ln (1-\eta )\right] 
\frac{1}{T^{\ast 2}}.
\end{eqnarray}%
by integrating with respect to density the {\it compressibility equation} %
\cite{Baxter71a,note3}

\begin{equation}
\left( \frac{\partial \beta P}{\partial \rho }\right) _{T}=\left[ 1-2\pi
\rho \ \widehat{q}\left( 0\right) \right] ^{2}=a^{2},  \label{eosc}
\end{equation}%
where $\widehat{q}\left( k\right) $ denotes the unidimensional Fourier
transform of $q(r)$.

On the other hand, no expressions for the SHS2-MSA virial and energy EOS
were given either in Ref. \cite{Mier89} or in subsequent literature on this
subject, to the best of our knowledge. We thus tackle this analysis here.

The most direct way we have followed is to consider the $z\rightarrow
+\infty $ limit of the HSY-MSA virial and energy pressures. Convenient
expressions for these quantities were given by Cummings and Smith (Eqs.
(18)-(20) of Ref. \cite{Cummings79a}). Unfortunately, their sticky limit
requires a rather elaborate analysis, which we will not report here.
However, the final result is rather simple: {\it both the virial and the
energy MSA pressures of the HSY fluid diverge in the sticky limit.} For the
MSA energy EOS, the same conclusion can be drawn by exploiting the
alternative form reported by Herrera {\it et al.} \cite{Herrera96}. It is
worth stressing again that this important feature has never been pointed out
before, to our knowledge.

To convince the reader that our statement is correct, we now provide a
simpler demonstration, based upon the analysis of the sticky limit of the
MSA second virial coefficients of the HSY fluid, as obtained from the
density expansion of $\left( Z^{{\rm HSY-MSA}}\right) _{C}$\ , $\left( Z^{%
{\rm HSY-MSA}}\right) _{V}$ and $\left( Z^{{\rm HSY-MSA}}\right) _{E}$. It
is well known that, in any approximate theory, compressibility, virial and
energy EOS's may yield different virial coefficients, and the first one at
which this difference begins to appear depends on the chosen closure. We now
show that, in the SHS2-MSA case, a non-consistency already appears at the $%
B_{2}$-level, and partly veils the singular character of the SHS2
Hamiltonian. Nonetheless, as it will be shown below, {\it both} $\left(
B_{2}^{{\rm HSY-MSA}}\right) _{V}$ {\it and} $\left( B_{2}^{{\rm HSY-MSA}%
}\right) _{E}$ are divergent, and this strongly supports the aforementioned
statement about the singular character of the virial and energy EOS for the
SHS2-MSA solution.

The quantity $\left( B_{2}^{{\rm HSY-MSA}}\right) _{C}$ can be easily
computed directly from the density expansion of $\partial \beta P/\partial
\rho $, given by Eq.(\ref{eosc}), upon using Eqs. (\ref{so2})-(\ref{so3}).
To evaluate $\left( B_{2}^{{\rm HSY-MSA}}\right) _{V}$ and $\left( B_{2}^{%
{\rm HSY-MSA}}\right) _{E}$ , it proves convenient to exploit the following
expressions for the MSA virial and energy EOS of \ a HSY fluid \cite%
{Hoye77b,Arrieta87}

\begin{equation}
Z_V^{{\rm HSY}}=1+4\eta \left[ g_{{\rm HSY}}(\sigma ^{+})-I\right] ,
\label{d6}
\end{equation}

\begin{equation}
Z_{E}^{{\rm HSY}}=Z_{{\rm HS}}+4\eta \left\{ \frac{1}{2}\left[ g_{{\rm HSY}%
}^{2}(\sigma ^{+})-g_{{\rm HS}}^{2}(\sigma ^{+})\right] -I\right\} ,
\label{d7}
\end{equation}

\begin{equation}
I=K \ \sigma ^{-3}\int_{\sigma }^{\infty }dr\ rg_{{\rm HSY}}(r)\ (1+zr)\
e^{-z(r-\sigma )}.  \label{d8}
\end{equation}%
Using the low-density expansion of the MSA approximation to $g_{{\rm HSY}%
}(r) $, Eq. ($\ref{clos5}$), one gets

\begin{equation}
\left( B_{2}^{{\rm HSY-MSA}}\right) _{C}=4v_{0}\left[ 1-\left( 1+\frac{1}{%
z\sigma}\right) \frac{1}{4 T ^{\ast }}\right] .  \label{d9}
\end{equation}

\begin{equation}
\left( B_{2}^{{\rm HSY-MSA}}\right) _{V}=4v_{0}\left[ 1-\left( 1+\frac{1}{%
z\sigma }\right) \frac{1}{4T^{\ast }}-\left( z\sigma +\frac{2}{3}z^{2}\sigma
^{2}\right) \frac{1}{192T^{\ast 2}}\right] .  \label{d10}
\end{equation}

\begin{equation}
\left( B_{2}^{{\rm HSY-MSA}}\right) _{E}=4v_{0}\left[ 1-\left( 1+\frac{1}{%
z\sigma }\right) \frac{1}{4T^{\ast }}-z\sigma \frac{1}{192T^{\ast 2}}\right]
.  \label{d11}
\end{equation}%
where $v_{0}=\left( \pi /6\right) \sigma ^{3}$ is the particle volume. The
difference among the three results for $B_{2}^{{\rm HSY-MSA}}$ is a clear
manifestation of the known thermodynamic inconsistency of the MSA \cite%
{Cummings79a,Cummings83}. It is however quite surprising that discrepancies
already appear at the $B_{2}$-level. Note that $\left( B_{2}^{{\rm HSY-MSA}%
}\right) _{C}>\left( B_{2}^{{\rm HSY-MSA}}\right) _{E}>\left( B_{2}^{{\rm %
HSY-MSA}}\right) _{V}\ $. These differences magnify with increasing $z$, and
become dramatic in the sticky limit which yields

\begin{equation}
\left( B_{2}^{{\rm SHS2-MSA}}\right) _{C}=4v_{0}\left( 1-\frac{1}{4T^{\ast }}%
\right) ,  \label{d12}
\end{equation}

\begin{equation}
\left( B_{2}^{{\rm SHS2-MSA}}\right) _{V}=-\infty =\left( B_{2}^{{\rm %
SHS2-MSA}}\right) _{E}\ .  \label{d13}
\end{equation}%
Eq. (\ref{d13}) thus confirms our previous statement about the divergence of
the MSA virial and energy pressures of the HSY fluid in the sticky limit. On
the other hand, Eq. (\ref{d12}) gives a finite value for $\left( B_{2}^{{\rm %
SHS2-MSA}}\right) _{C}$ in agreement with the corresponding compressibility
EOS. However, we stress that this result is due to the fact that the MSA
involves an approximation at the level of Boltzmann factor \cite{Arrieta91},
i.e. 
\begin{equation}
e_{{\rm MSA}}\left( r\right) =\theta (r-\sigma )\left[ 1-\beta \phi _{{\rm %
tail}}\left( r\right) \right] ,  \label{d14}
\end{equation}

\noindent equivalent to the linearization $\exp \left[ -\beta \phi _{{\rm %
tail}}\left( r\right) \right] \simeq 1-\beta \phi _{{\rm tail}}\left(
r\right) $ (as can be inferred from a comparison between the zeroth-order
terms in the density expansion of $g_{{\rm exact}}(r)$ and $g_{{\rm MSA}%
}\left( r\right) $, i.e. $g_{{\rm exact}}^{(0)}\left( r\right) =e(r)=e_{{\rm %
HS}}(r)\exp \left[ -\beta \phi _{{\rm tail}}\left( r\right) \right] $ and $%
g_{{\rm MSA}}^{(0)}\left( r\right) =e_{{\rm HS}}(r)\left[ 1-\beta \phi _{%
{\rm tail}}\left( r\right) \right] $, respectively). The replacement of the
exact $e_{{\rm HSY}}(r)$ in the expression for $B_{2}^{{\rm HSY}}$ with its
MSA counterpart (\ref{d14}) just leads to the $\left( B_{2}^{{\rm SHS2-MSA}%
}\right) _{C}$ \ result, in the sticky limit.

Summarizing, we can conclude that the existence of the MSA solution, $q_{%
{\rm SHS2-MSA}}(r)$, is a fortuitous consequence of the fact that the
divergent character of the SHS2 model is {\it masked} by the MSA
linearization ($\ref{d14}$).

\subsection{MSA internal energy}

As a final point of this first part, we wish to point out an interesting
different behaviour between the internal energy in the SHS2-MSA and SHS1-PY
solution.

Let us now consider the {\it energy }$(E)$ {\it equation,}

\begin{equation}
u_{{\rm ex}}=U_{{\rm ex}}/N=2\pi \rho \int_{0}^{\infty }dr\ r^{2}g(r)\phi
(r),  \label{eose}
\end{equation}%
where $U_{{\rm ex}}$ is the excess internal energy. Within the MSA, one gets
for our HSY fluid

\begin{eqnarray*}
-\beta u_{{\rm ex}}^{{\rm HSY-MSA}} &=&2\pi \rho \int_{\sigma }^{\infty }dr\
r^{2}\left[ 1+\gamma _{{\rm HSY-MSA}}\left( r\right) -\beta \phi _{{\rm Y}%
}\left( r\right) \right] \left[ -\beta \phi _{{\rm Y}}\left( r\right) \right]
\\
&=&2\pi \rho \left\{ K_{0}\int_{\sigma }^{\infty }dr\ r\left[ 1+\gamma _{%
{\rm HSY-MSA}}\left( r\right) \right] \ ze^{-z(r-\sigma )}+\frac{K_{0}^{2}}{2%
}z\right\} .
\end{eqnarray*}%
Since $\gamma _{{\rm HSY-MSA}}(r)$ is finite and sufficiently regular
everywhere, it is easy to show, integrating by parts or using the property $%
\lim_{z\rightarrow +\infty }ze^{-z(r-\sigma )}\theta (r-\sigma )=\ \delta
_{+}(r-\sigma )$ \cite{note2}, \ that the last integral remains finite as $%
z\rightarrow +\infty $. \noindent\ In conclusion, $u_{{\rm ex}}^{{\rm HSY-MSA%
}}$ {\it diverges linearly}. This should be compared with the {\it %
logarithmic }divergence found in SHS1-PY energy \cite{Baxter68}. We
speculate that this difference in the internal energy behaviour parallels
that in the corresponding EOS. It is then clear that, in the SHS1-PY case,
such a weak divergence does not in itself constitute an impediment to the
existence of finite EOS's in the sticky limit \cite%
{Baxter68,Stell91,Yuste93a}.

As a side comment, it is worth noticing that our finding contradicts Ref. %
\cite{Herrera98}, where an MSA energy with a non-singular sticky
contribution (Eq. (42) of that paper) is reported .

\section{NEW\ MODEL AND\ NEW\ CLOSURE}

Given the above premise, one could, at this point, suspect that all previous
findings based upon the SHS2 model (and notably, those referring to
polydisperse colloidal fluids \cite{Gazzillo00,Gazzillo02a,Gazzillo02b})
should be discarded. This is not so, and all those results are, in fact,
correct. The second aim of our paper is to show that an analytic solution
with {the same functional form as $q_{{\rm SHS2-MSA}}(r)$ is obtained within
a new and well-defined model (which we call SHS3), coupled with a simple new
closure. As a consequence, most of the results derived from SHS2 can be
recovered after an appropriate re-interpretation. }

\subsection{SHS3 potential}

We define a new SHS Hamiltonian, which represents the simplest correct
alternative to SHS2, and is analytically solvable within a novel closure.
The basic idea hinges on Baxter's trick of a logarithmic tail, combined with
the infinite-ranged Yukawa expression of SHS2. The starting potential
(hereafter referred to as M3) is

\begin{equation}
\beta \phi _{{\rm M3}}(r)=\left\{ 
\begin{array}{ccc}
+\infty , &  & 0<r<\sigma , \\ 
-\ln \left[ 1+zK_{0}\ e^{-z(r-\sigma )}/r\right] , &  & r\geq \sigma ,%
\end{array}%
\right.  \label{m1}
\end{equation}%
where $K_{0}$ depends on $T$ according to Eq. ($\ref{i2}$), and the sticky
limit corresponds again to $z\rightarrow +\infty .$

We note that at large $r$ values, the quantity $x\equiv -\beta \phi _{{\rm Y}%
}\left( r\right) =zK_{0}\ e^{-z(r-\sigma )}/r$, appearing in the argument of
the logarithm, becomes so small that the approximation $\ln (1+x)\simeq x$
can be used. In other words, M3 has the same large-$r$ asymptotic behaviour
as $-\beta \phi _{{\rm Y}}\left( r\right) $, but differs from the Yukawa
tail near contact. This is depicted in Figure 1, where it can be clearly
seen that this difference between the two potentials increases as $z$
increases. However, for the M3 model, the Boltzmann factor reads%
\begin{equation}
e_{{\rm M3}}(r)=\theta \left( r-\sigma \right) \left[ 1+zK_{0}\
e^{-z(r-\sigma )}/r\right] ,  \label{m2}
\end{equation}%
with the essential consequence that the corresponding {\it exact }second
virial coefficient remains finite in the sticky limit, i.e.,

\begin{equation}
B_{2}^{{\rm M3-exact}}=4v_{0}\left[ 1-3\varepsilon _{0}^{\ast }\left( 1+%
\frac{1}{z\sigma }\right) \right] ,
\end{equation}

\begin{equation}
B_{2}^{{\rm SHS3-exact}}=\lim_{z\rightarrow +\infty }B_{2}^{{\rm M3-exact}%
}=4v_{0}\left( 1-\frac{1}{4T^{\ast }}\right) .  \label{b2}
\end{equation}%
Note that this $B_{2}$-expression, {\it exact }for SHS3, is formally
identical with $B_{2}^{{\rm SHS2-MSA}}$, which is only {\it approximate} for
SHS2.

As a further check that we are on the right track, we have also computed the 
{\it exact }SHS3 third virial coefficient

\begin{equation}
B_{3}^{{\rm SHS3-exact}}=\lim_{z\rightarrow +\infty }B_{3}^{{\rm M3-exact}%
}=v_{0}^{2}\left( 10-\frac{5}{T^{\ast }}+\frac{1}{T^{\ast 2}}-\frac{1}{%
18T^{\ast 3}}\right) .  \label{b3}
\end{equation}%
It is remarkable that $B_{3}^{{\rm SHS3-exact}}$ coincides with $B_{3}^{{\rm %
SHS1-exact}}$, once $T^{\ast }$ is replaced with its counterpart $\tau $ of
Baxter's model \cite{Baxter68}.

\subsection{Modified MSA closure and solution}

We define a {\it modified mean spherical approximation} (mMSA)

\begin{equation}
c_{{\rm mMSA}}\left( r\right) =\left\{ 
\begin{array}{ccc}
-\left[ 1+\gamma \left( r\right) \right] , &  & 0<r<\sigma , \\ 
f_{{\rm tail}}\left( r\right) =\exp \left[ -\beta \phi _{{\rm tail}}\left(
r\right) \right] -1, &  & r\geq \sigma .%
\end{array}%
\right.  \label{clos3}
\end{equation}%
In terms of \ RDF, this closure reads

\begin{eqnarray}
g_{{\rm mMSA}}\left( r\right) &=&e_{{\rm HS}}\left( r\right) \left[ 1+\gamma
\left( r\right) +f_{{\rm tail}}\left( r\right) \right]  \nonumber \\
&=&e\left( r\right) +e_{{\rm HS}}\left( r\right) \gamma \left( r\right) ,
\label{clos4}
\end{eqnarray}%
since $e_{{\rm HS}}\left( r\right) \left[ 1+f_{{\rm tail}}\left( r\right) %
\right] =e\left( r\right) $.

Although this closure is not completely new (see Ref.s \cite{Huang84,Pini02}%
), it was never formulated in the present general form. Yet it is a very
natural choice, since it yields the correct zeroth-order term in the density
expansion of the DCF: $c_{{\rm exact}}(r)$ $\rightarrow $ $f(r)$ when $\rho
\rightarrow 0.$

The advantage of coupling the M3 potential with the mMSA is that one finds $%
c_{{\rm M3-mMSA}}(r)=f\left( r\right) =zK_{0}\ e^{-z(r-\sigma )}/r$ ($r\geq
\sigma $), which is identical with the usual Yukawa closure. This mapping
allows the immediate identification between the solution $q_{{\rm M3-mMSA}%
}(r)$ with $q_{{\rm HSY-MSA}}(r)$, and thus in the limit $z\rightarrow
+\infty $ we get

\begin{equation}
q_{{\rm SHS3-mMSA}}(r)=q_{{\rm SHS2-MSA}}(r).
\end{equation}%
In short, the solution given by Eqs. ($\ref{so1}$)-($\ref{so3}$) is
recovered, but within a new {\it well-defined} model, with a {\it finite}
second virial coefficient.

\subsection{Equations of state}

\subsubsection{Compressibility route}

Since $Z_{C}^{{\rm SHS3-mMSA}}$ can be computed directly from $q_{{\rm %
SHS3-mMSA}}(r)$ and Eq. (\ref{eosc}) (in fact, sticky limit and $\rho $%
-integration commutate), it results that

\begin{equation}
Z_{C}^{{\rm SHS3-mMSA}}=Z_{C}^{{\rm SHS2-MSA}}.  \label{zc}
\end{equation}

Expanding this quantity in powers of $\rho $, one finds that $\left( B_{2}^{%
{\rm SHS3-mMSA}}\right) _{C}$ $=$ $B_{2}^{{\rm SHS3-exact}}$, whereas $%
\left( B_{3}^{{\rm SHS3-mMSA}}\right) _{C}=v_{0}^{2}\left[ 10-\frac{10}{3}%
\left( T^{\ast }\right) ^{-1}+\frac{1}{3}\left( T^{\ast }\right) ^{-2}\right]
$ differs from $B_{3}^{{\rm SHS3-exact}}$.

\subsubsection{Virial route}

Let us insert Eq. ($\ref{clos4}$) into the {\it virial }$(V)$ {\it equation,}

\begin{equation}
Z_{V}=1+\frac{2\pi }{3}\rho \int_{0}^{\infty }dr\ r^{3}g(r)\left[ -\beta
\phi ^{\prime }(r)\right] \equiv 1+\frac{2\pi }{3}\rho J_{V}.  \label{eosv}
\end{equation}%
The integral can be written as $J_{V}=J_{1}+J_{2}$, \ with

\[
J_{1}=\int_{0}^{\infty }dr\ r^{3}e(r)\left[ -\beta \phi ^{\prime }(r)\right]
=\int_{0}^{\infty }dr\ r^{3}e^{\prime }(r)=\int_{0}^{\infty }dr\
r^{3}f^{\prime }(r), 
\]%
\begin{eqnarray*}
J_{2} &=&\int_{0}^{\infty }dr\ e_{{\rm HS}}\left( r\right) r^{3}\gamma
\left( r\right) \left[ -\beta \phi ^{\prime }(r)\right] \\
&=&\int_{0}^{\infty }dr\ e_{{\rm HS}}\left( r\right) r^{3}\gamma \left(
r\right) \left[ -\beta \phi _{{\rm HS}}^{\prime }(r)-\beta \phi _{{\rm tail}%
}^{\prime }(r)\right] .
\end{eqnarray*}%
Integrating $J_{1}$ by parts, and observing that the boundary terms vanish,
one finds that $\left( 2\pi /3\right) J_{1}=B_{2}$. On the other hand, if $%
J_{2}$ is written as $J_{2a}+J_{2b}$, with $J_{2a}=\int_{0}^{\infty }dr\ e_{%
{\rm HS}}\left( r\right) r^{3}\gamma \left( r\right) \left[ -\beta \phi _{%
{\rm HS}}^{\prime }(r)\right] $, and using $e_{{\rm HS}}\left( r\right) %
\left[ -\beta \phi _{{\rm HS}}^{\prime }(r)\right] =e_{{\rm HS}}^{\prime
}\left( r\right) =\delta (r-\sigma )$, one gets%
\begin{equation}
Z_{V}^{{\rm mMSA}}=1+B_{2}\rho +\frac{2\pi }{3}\rho \left\{ \sigma
^{3}\gamma _{{\rm mMSA}}\left( \sigma \right) +\int_{\sigma }^{\infty }dr\
r^{3}\gamma _{{\rm mMSA}}\left( r\right) \left[ -\beta \phi _{{\rm tail}%
}^{\prime }(r)\right] \right\} .  \label{v1}
\end{equation}

In the particular case of the M3 potential, $-\beta \phi _{{\rm tail}%
}^{\prime }(r)$ assumes increasingly large negative values as $z$ increases,
whereas $\gamma _{{\rm M3-mMSA}}(r;z)\leq \gamma _{{\rm M3-mMSA}}(\sigma ;z)$
remains bounded, in such a way that the integral $J_{2b}$ diverges in the
sticky limit (for shortness, our detailed analysis of this point is not
reported here).

Eq. ($\ref{v1}$) and these results imply that the mMSA-virial EOS diverges.
However, since the second virial coefficient is finite and exact, and $%
\left( B_{3}^{{\rm SHS3-mMSA}}\right) _{V\text{ }}$ diverges whereas $B_{3}^{%
{\rm SHS3-exact}}$ is finite, the singularity of $Z_{V}^{{\rm SHS3-mMSA}}$
is not due to the Hamiltonian, but is surely caused by the mMSA closure.

\subsubsection{Energy route}

Finally let us explore the energy (or free energy) route for
temperature-dependent potentials. It can be shown \cite{Baxter71b} that, for
any potential,

\begin{equation}
\frac{\partial }{\partial \zeta }\left[ \frac{\beta \left( A-A_{{\rm id}%
}\right) }{N}\right] =2\pi \rho \int_{0}^{\infty }dr\ r^{2}g(r)\ \frac{%
\partial \left[ \beta \phi \left( r\right) \right] }{\partial \zeta },
\label{e0}
\end{equation}%
where $\zeta $ denotes an arbitrary parameter from which $\beta \phi \left(
r\right) $ depends upon, and $A_{{\rm id}}$ indicates the ideal gas free
energy. In particular, when $\zeta =\beta $ and $g(r)=g_{{\rm mMSA}}\left(
r\right) $, one gets%
\[
\frac{\partial }{\partial \beta }\left[ \frac{\beta \left( A-A_{{\rm id}%
}\right) }{N}\right] _{{\rm mMSA}}=2\pi \rho \left( -\int_{0}^{\infty }dr\
r^{2}\frac{\partial e\left( r\right) }{\partial \beta }+\int_{\sigma
}^{\infty }dr\ r^{2}\gamma _{{\rm mMSA}}\left( r\right) \frac{\partial \left[
\beta \phi _{{\rm tail}}\left( r\right) \right] }{\partial \beta }\right) . 
\]%
For the M3 potential, this expression becomes

\begin{equation}
\frac{\partial }{\partial \beta }\left[ \frac{\beta \left( A_{{\rm M3-mMSA}%
}-A_{{\rm id}}\right) }{N}\right] =-12\eta \ \varepsilon _{0}\left[ 1+\frac{1%
}{z\sigma }+\frac{\gamma _{{\rm M3-mMSA}}\left( \sigma ;\beta ,z\right) }{%
1+z\sigma \ \beta \varepsilon _{0}}+{\cal O}\left( z^{-2}\right) \right] ,
\label{e0b}
\end{equation}%
With the boundary condition $T\rightarrow \infty $ ($\beta \rightarrow 0$),
corresponding to the HS case, this equation can be integrated with respect
to $\beta ,$ and in the $z\rightarrow +\infty $ limit one finds%
\begin{equation}
\beta (A_{{\rm SHS3-mMSA}}-A_{{\rm HS}})/N=-\frac{1}{T^{\ast }}\eta ,
\label{e2}
\end{equation}%
(since sticky limit and $\beta $-integration commutate). From $%
Z=\eta \partial (\beta A/N)/\partial \eta $ we then get

\begin{equation}
Z_{E}^{{\rm SHS3-mMSA}}=Z_{{\rm HS}}-\frac{1}{T^{\ast }}\eta .  \label{ze}
\end{equation}

It is noteworthy that these results for the Helmholtz free energy and the
energy-EOS are van der Waals-like. The effect of the surface adhesion enters
these expressions only at the level of the second virial coefficient (which
is exact), while all higher-order virial coefficients predicted by Eq. ($\ref%
{ze}$) coincide with the pure HS ones. The reason for this may be traced
back to the fact that the mMSA closure takes into account only the
zeroth-order term in the density expansion of the DCF outside the core.

\section{CONCLUSIVE REMARKS}

Three different models of `sticky hard spheres' have been treated in this
paper. The first two (SHS1 and SHS2) were already present in the literature,
while the last one (SHS3) is new and constitutes the main contribution of
the present work.

All these models describe a fluid of rigid spherical particles with
infinitely strong surface adhesion, defined through an appropriate `sticky
limit' which constitutes an essential part of the model. The choice of the
starting potential strongly influences that of the approximate closure to be
used to solve the OZ equation analytically. In SHS1 a suitably defined
square-well allows the solution within the PY approximation, whereas in SHS2
the starting point is a HS-Yukawa potential which requires the MSA
approximation. However, the most crucial point is that, while SHS1 is a
perfectly well-defined model, SHS2 is {\it not,} since its {\it exact}
second virial coefficient is divergent, as we have proven in this paper.
This singularity is mirrored by a similar behaviour appearing in both the
virial and energy SHS2-MSA EOS's, but it should be emphasized that it is a
weakness of the model itself and is not due to the MSA approximation. This
means that the SHS2 model itself is not properly defined and must be
abandoned.

To replace it, a proposal has been put forward in the second part of the
paper. Our recipe is based upon the introduction of a new potential (M3),
coupled with a new closure (mMSA), in such a way that the Baxter-OZ
equations turn out to be analytically solvable, with formally the same
solution $q(r)$ found in the SHS2-MSA case. This remarkable correspondence
allows a recovery of all previous results for structure factors and
compressibility EOS based upon the SHS2-MSA solution \cite%
{Tutschka98,Gazzillo00,Gazzillo02a,Gazzillo02b}, after a re-interpretation
in terms of SHS3-mMSA.

The SHS3 model is the simplest correct alternative to SHS2. It is
well-defined, since the sticky limit of the {\it exact} second virial
coefficient, $B_{2\text{ }}^{{\rm SHS3-exact}}$, remains {\it finite. }%
Furthermore, its starting potential, M3, is asymptotically equivalent to the
HSY one at large $r$ values, being different from it only in the contact
region.

As regards the closure, the mMSA is correct in the zero-density limit, and
has, consequently, a higher thermodynamic consistency than the MSA. As a
matter of fact, compressibility, virial and energy routes all generate the
same, exact, $B_{2}$. On the other hand, one cannot expect from the mMSA to
go further in\ consistency, since this closure is still rather poor, taking
into account only the zeroth-order term in the density expansion of the DCF
outside the core. We have shown that discrepancies are already found at the
level of third virial coefficient: $\left( B_{3\text{ }}^{{\rm SHS3-mMSA}%
}\right) _{C}$ is temperature-dependent but different from $B_{3\text{ }}^{%
{\rm SHS3-exact}}$, $\left( B_{3\text{ }}^{{\rm SHS3-mMSA}}\right) _{V}$
diverges, while $\left( B_{3\text{ }}^{{\rm SHS3-mMSA}}\right) _{E}$ is
temperature-independent and equal to its HS counterpart. Clearly, these
three results could coincide within a more refined (density-dependent)
closure.

It would be very interesting to investigate whether the SHS3 model admits
further analytic solutions, within more sophisticated approximations (such
as, in particular, the PY approximation). We plan to do such analysis in
future work, including also a more detailed comparison with Baxter's
original model.

Finally, it is worth noting that the extension of the SHS3 model to mixtures
can be easily carried out. We hope that the corresponding solution can
provide a simple useful tool for further studies on structural and
thermodynamic properties of polydisperse colloidal fluids.

\vskip 1cm

\acknowledgments This work was partially supported by the Italian INFM
(Istituto Nazionale per la Fisica della Materia). Useful discussions with
Lesser Blum, Gerhard Kahl and Yurii Kalyuzhnyi are gratefully acknowledged.
A special thank goes to Andr\'{e}s Santos for a suggestion concerning the
virial EOS.

\bigskip 

\newpage

\begin{figure}[tbp]
\centerline{ \epsfxsize=7.0truein \epsfysize=5.0truein
\epsffile{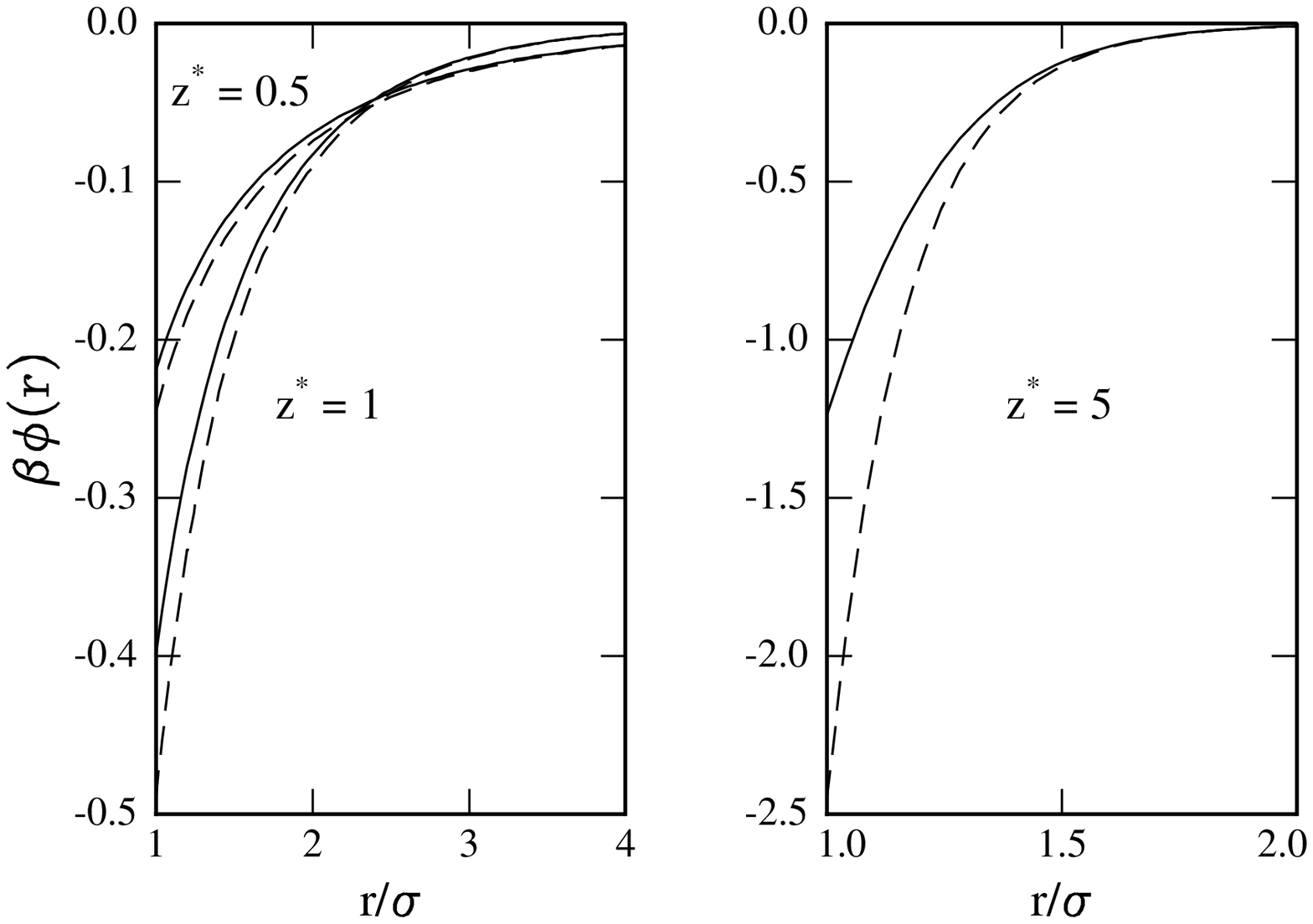} }
\vskip1.0cm
\caption{Comparison between $\protect\beta \protect\phi _{{\rm M3}}(r)$
(solid curves) and $\protect\beta \protect\phi _{{\rm HSY}}(r)$ (dashed
curves) at $T^{*}=0.17$, for increasing values of the dimensionless
inverse-range parameter, $z^{*}=z \protect\sigma$. Note the difference in
both vertical and horizontal scales between the two parts of the figure.}
\label{f:Fig1}
\end{figure}


\begin{references}
\bibitem{note3} In the case of mixtures, additional requirements are
necessary to perform this operation (see \cite{Gazzillo02b}).
\bibitem{note2} Here, $\ \delta _{+}(x)$ is the asymmetrical Dirac delta
function defined by: $\int_{A}^{B}dx\ F(x)\delta _{+}(x-x_{0})=F\left(
x_{0}^{+}\right) ,$ if $A\leq x_{0}<B$, and \ $=0,$ if $x_{0}<A$ or $%
x_{0}\geq B$ (the Yukawa tail, $\phi _{{\rm Y}}(r)=-z\varepsilon _{0}\sigma
^{2}\ r^{-1}e^{-z(r-\sigma )}\theta (r-\sigma ),$\ gives rise to a
singularity, since $\lim_{z\rightarrow +\infty }ze^{-z(r-\sigma )}\theta
(r-\sigma )=\ \delta _{+}(r-\sigma )$\cite{Ginoza96}).
\end{references}
\end{document}